\documentclass[doublecol]{epl2}
\usepackage{graphicx}  
\usepackage{latexsym}  
\font\fiverm=cmr5
\input prepictex
\input pictex
\input postpictex

\def\mathput#1{\relax \ifmmode \displaystyle #1\else $\displaystyle #1$\fi}
\def\beq{\begin{equation}}
\def\eeq{\end{equation}}

\title{Light scattering by radiation fields: the optical medium analogy}
\shorttitle{Light scattering by radiation fields}

\author{D. Bini\inst{1,2,3,4} \and P. Fortini\inst{5} \and A. Geralico\inst{6} \and M. Haney\inst{2,6} \and A. Ortolan\inst{7}}

\institute{
\inst{1} Istituto per le Applicazioni del Calcolo \lq\lq M. Picone," CNR, I--00185 Rome, Italy\\
\inst{2} ICRA, \lq\lq Sapienza" University of Rome, I--00185 Rome, Italy \\
\inst{3} INFN sezione di Firenze, I--00185 Sesto Fiorentino (FI), Italy \\
\inst{4} Astronomical Observatory of Torino, INAF, via Osservatorio 20, I-10025 Pino Torinese (TO), Italy \\
\inst{5} Department of Physics, University of Ferrara and INFN Sezione di Ferrara, I--44100 Ferrara, Italy \\
\inst{6} Physics Department, \lq\lq Sapienza" University of Rome, I--00185 Rome, Italy \\
\inst{7} INFN - National Laboratories of Legnaro, I--35020 Legnaro (PD), Italy
 } 


\abstract{
The optical medium analogy of a radiation field generated by either an exact gravitational plane wave or an exact electromagnetic wave {in the framework of general relativity} is developed.
The equivalent medium of the associated background field is inhomogeneous and anisotropic in the former case, whereas  it is inhomogeneous but isotropic {in the latter}.
The features of light scattering are investigated by assuming the interaction region to be sandwiched between two flat spacetime regions, where light rays propagate along straight lines.
Standard tools of ordinary wave optics are used to study the deflection of photon paths due to the interaction with the radiation fields, {allowing for} a comparison between the optical properties of the equivalent media associated with the different background fields.
}

\begin{document}

\maketitle

\section{Introduction}

An electromagnetic field in any gravitational background, associated with the metric $ds^2=g_{\alpha\beta}dx^\alpha dx^\beta$ and Cartesian-like coordinates $(t,x,y,z)$, can be thought of as propagating in flat spacetime but in the presence of a medium whose properties are determined by conformally invariant quantities constructed from the metric tensor. 
In fact, the covariant electromagnetic field $F_{\mu\nu}$ and its rescaled contravariant counterpart $\sqrt{-g}F^{\mu\nu}$ can be decomposed as $F_{\mu\nu}\to ({\mathbf E},{\mathbf B})$ and $\sqrt{-g}F^{\mu\nu}\to (-{\mathbf D},{\mathbf H})$ to yield the usual Maxwell's equations in a medium \cite{pleb,skrotskii,fdf1,mas73}
\begin{eqnarray}
\label{MWEQ}
&& \nabla \cdot {\mathbf B}=0\,,\quad \nabla \times {\mathbf E}=-\partial_t {\mathbf B}\,,\nonumber\\
&& \nabla \cdot {\mathbf D}=4\pi\rho\,,\quad \nabla \times {\mathbf H}=\partial_t {\mathbf D}+4\pi {\mathbf J}\,,
\end{eqnarray}
where $J_\mu \to (\rho , {\mathbf J})$ satisfies the conservation law
\beq
\partial_t \rho+ \nabla \cdot {\mathbf J}=0\,.
\eeq
To these equations the constitutive relations
\beq
\label{diel}
D_a=\epsilon_{ab}E_b -({\mathbf M}\times {\mathbf H})_a\,,\qquad 
B_a=\mu_{ab}H_b+({\mathbf M}\times {\mathbf E})_a\,,
\eeq
{must be added}, 
where\footnote{We use here geometrical units so that $c=1=G$ and $-+++$ for the Lorentzian signature of the metric tensor. Greek indices run from $0$ to $4$ {and} latin ones from $1$ to $3$.
}
\beq
\label{epsdef}
\epsilon_{ab}=\mu_{ab}=-\sqrt{-g}\, \frac{g^{ab}}{g_{tt}}\, 
\eeq
play the role of  electric and magnetic permeability tensors and 
\beq
M_a=-\frac{g_{ta}}{g_{tt}}\
\eeq
is a vector field {associated with} rotations of the reference frame.
Within this framework, the description of electromagnetic fields in a curved spacetime is equivalently accomplished by solving Maxwell's equations in a flat spacetime but in the presence of a medium, whose properties are fully specified by the associated constitutive relations.
Such a material medium is in general anisotropic and has no birefringence, due to the proportionality between polarization tensors.
Notice that the conformal invariance of Maxwell's equations is reflected, in the present formulation, by the independence of the dielectric tensors $\epsilon_{ab}$ and $\mu_{ab}$ from a conformal factor in the metric components, a property also shared by the \lq\lq spatial" vector ${\mathbf M}$.

Following \cite{mas73}, if one introduces the  complex vectors
\beq
\label{FpmSpm}
{{\mathbf F}} ={{\mathbf E}}+ i {{\mathbf H}}\,,\qquad 
{{\mathbf  S}} ={{\mathbf  D}}+ i {{\mathbf  B}}\,,
\eeq
the constitutive relations read
\beq
S_a=\epsilon_{ab}F_b+ i ({\mathbf M}\times {\mathbf F})_a\,.
\eeq
Furthermore, it is possible to write the electromagnetic equations (\ref{MWEQ}) in a form which is particularly suitable for the discussion of wave phenomena, i.e.,
\beq
\frac{1}{i}\nabla \times {\mathbf F} =\partial_t {\mathbf S} + 4\pi {\mathbf J}\,,\qquad 
\nabla \cdot {\mathbf S}=4\pi\rho\,.
\eeq
Plane waves satisfy the above equations in a small spacetime region, where the metric tensor can be always assumed to vary in space and time only slightly with respect to the wavelength and the period of the wave, respectively.
Therefore, one can apply there the standard tools of ordinary wave optics to study the optical properties of the equivalent medium.
For instance, in {the} absence of currents, one may look for solutions of the form
\beq
\label{toreplace}
{\mathbf E}={\mathbf E}_0 e^{i(n {\mathbf k}\cdot {\mathbf x}-\omega t)}\,,\quad  {\mathbf E}_0=const.\,,\quad |{\mathbf k}|=\omega\,,
\eeq
where $n$ is an effective refraction index.
Similar expressions hold for ${\mathbf B}$, ${\mathbf H}$ and ${\mathbf D}$. 
Maxwell's equations (\ref{MWEQ}) then imply
\beq
\label{MWEQsol}
{\mathbf n}\times {\mathbf E}={\mathbf B} \,,\qquad {\mathbf n}\times {\mathbf H}=-{\mathbf D}
\eeq
and ${\mathbf n}\cdot {\mathbf B}=0={\mathbf n}\cdot {\mathbf D}$, where ${\mathbf n}=(n/\omega){\mathbf k}=n {\mathbf e}$ and ${\mathbf e}={\mathbf k}/\omega$ is the spatial unit vector of the photon direction. 
{A substitution of} the constitutive relations (\ref{diel}) into Eqs. (\ref{MWEQsol}) leads to
\beq
[\epsilon_{ik}+\varepsilon_{ijr}\varepsilon_{suk}({\mathbf n}-{\mathbf M})_j({\mathbf n}-{\mathbf M})_u (\epsilon^{-1})_{rs}]E_k=0\,,
\eeq
where $\varepsilon_{ijk}$ denotes the Levi-Civita alternating symbol. {One finds a similar equation for $H_k$.}
The existence of  eigensolutions for ${\mathbf E}$ implies the generalized Fresnel equation
\beq
n^2\epsilon_{{\mathbf e}{\mathbf e}} -2n\epsilon_{{\mathbf e}{\mathbf M}} +\epsilon_{{\mathbf M}{\mathbf M}}-{\rm det}(\epsilon)=0\,,
\eeq
where the compact notation $X_{ab}A^aB^b=X_{{\mathbf A}{\mathbf B}}$ has been introduced for contraction of a generic matrix $X_{ab}$ with vectors $A_a$ and $B_b$.
The above equation gives the relation between {the effective refraction index} of the medium and the components of the polarization tensors and the direction of propagation of the electromagnetic wave \cite{skrotskii}.
The solution for the refraction index is thus given by 
\beq
n=\frac{1}{\epsilon_{{\mathbf e}{\mathbf e}}}\left[\epsilon_{{\mathbf e}{\mathbf M}}
+\sqrt{{\rm det}(\epsilon)[\epsilon_{{\mathbf e}{\mathbf e}}-(\epsilon^{-1})_{{\mathbf c}{\mathbf c}}]}\right]\,,
\eeq
where ${\mathbf c}={\mathbf M}\times {\mathbf e}$ and ${\rm det}(\epsilon)=\epsilon_1\epsilon_2\epsilon_3$, in a coordinate system in which electric and magnetic permeability tensors are diagonal, i.e., $\epsilon_{ab}={\rm diag}[\epsilon_1,\epsilon_2,\epsilon_3]=\mu_{ab}$.
In the special case $M_a=0$, the above expression simplifies to
\beq
\label{ndefstat}
n =\sqrt{\frac{{\rm det}(\epsilon)}{\epsilon_{{\mathbf e}{\mathbf e}}}}\,.
\eeq

In the literature the \lq\lq optical medium analogy" has been widely used to study both the propagation of electromagnetic waves in a gravitational field and scattering processes by either static or rotating  black holes as well as by some cosmological solution, e.g., the G\"odel universe \cite{fdf1,mas73,mas74,mas75}. Mashhoon and Grishchuk \cite{masgris80} have also considered some weak gravitational field background associated with a gravitational plane wave.
In the present work we are interested in investigating the scattering of light by a radiation field given by either an exact gravitational plane wave or an exact electromagnetic wave.
We assume that the interaction region is sandwiched between two flat spacetime regions.
A similar situation has been considered in Ref. \cite{wavesPR}, where the scattering of massive particles was studied instead. 
We first develop the equivalent medium analogy of the associated background fields, {by} exploring the optical properties of the corresponding effective material media. 
Using standard tools of ordinary wave optics we then study the deflection of photon paths due to the interaction with the radiation field, {allowing for} a comparison between the optical properties of the equivalent media associated with the different background fields.

\section{Radiation scattering from a sandwiched gravitational wave}

The background of a radiation field (either a gravitational plane wave or an electromagnetic wave) propagating along the $z$-axis of an adapted coordinate system $(t,x,y,z)$ is described by the line element \cite{grif2}
\begin{eqnarray}
\label{lineelement}
ds^2&=&-dt^2+dz^2+ F(t-z)^2dx^2+G^2(t-z)dy^2\nonumber\\
&=&-2dudv+F(u)^2dx^2+G^2(u)dy^2\,.
\end{eqnarray}
where the null coordinates $u=(t-z)/\sqrt{2}$ and $v=(t+z)/\sqrt{2}$ are related to the Cartesian ones in a standard way. 

In this section we consider the (vacuum) case of a gravitational plane wave with a single polarization state ($+$ state), corresponding to metric functions
\beq
F(u)=\cos b_{\rm(gw)} u\,,\qquad G(u)=\cosh b_{\rm(gw)} u\,,
\eeq
where the background quantity $b_{\rm(gw)}$ is related to the frequency of the wave by $b_{\rm(gw)}=\sqrt{2}\omega_{\rm(gw)}$.

Let the plane gravitational wave be sandwiched between two Minkowskian regions $u\in (-\infty, 0)\cup (u_1, \infty)$.
{The coordinate horizon of the metric (\ref{lineelement}) at $b_{\rm (gw)}u=\frac{\pi}{2}$,  where it  is degenerate, is avoided by} 
restricting the coordinate $u$ to the interval $[0,u_1]$ with $b_{\rm (gw)}u_1<\frac{\pi}{2}$. 
The matching conditions at the boundaries of the sandwich impose restrictions on the metric functions $F$ and $G$ before and after the passage of the wave where the spacetime is Minkowskian. A possible choice to extend the metric to all values of $u$ is the {following \cite{rindler}}
\begin{eqnarray}
\label{Eq8}
F(u)&=&\left\{
  \begin{array}{llc} 1  & u\le0 & \hbox{\rm(I) } \\
                   \cos b_{\rm(gw)} u &  0\le u\le u_1  & \hbox{\rm(II) }    \\
                   \alpha+\beta u & u\ge u_1   &  \hbox{\rm(III) }
  \end{array}\right. \nonumber
\\
G(u)&=&\left\{
  \begin{array}{llc} 1   & u\le 0  & \hbox{\rm(I) }  \\
                   \cosh b_{\rm(gw)}u&  0\le u\le u_1 & \hbox{\rm(II) }    \\
                   \gamma+\delta u, &  u\ge u_1   &  \hbox{\rm(III) }
  \end{array}\right.
\end{eqnarray}
where labels I, II and III refer to in-zone ($u\le 0$), wave-zone ($0<u<u_1$) and out-zone $(u\ge u_1)$, respectively. The constants $\alpha$, $\beta$, $\gamma$ and $\delta$ can be found by requiring $C^1$ regularity conditions for the metric functions at the boundaries $u=0$ and $u=u_1$ of the sandwich, namely
\begin{eqnarray}
\alpha &=& \cos b_{\rm(gw)} u_1 +b_{\rm(gw)} u_1 \sin b_{\rm(gw)} u_1\,, \nonumber\\
\beta& = &-b_{\rm(gw)} \sin b_{\rm(gw)} u_1\,,\nonumber \\
\gamma&=& \cosh b_{\rm(gw)} u_1 -b_{\rm(gw)} u_1 \sinh b_{\rm(gw)} u_1\,, \nonumber\\
\delta& =& b_{\rm(gw)} \sinh b_{\rm(gw)} u_1\,.
\end{eqnarray}

\begin{figure}[t]
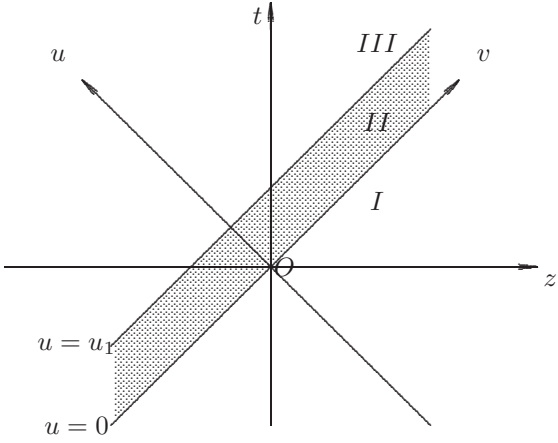

\label{fig:1}
$$ \vbox{
\beginpicture
  \setcoordinatesystem units <0.35cm,0.35cm> point at 0 0
\putrule from -10 0 to  10 0
\putrule from 0 -6 to  0 10

\arrow <.25cm> [.15,.25]    from  0 9  to 0 10
\arrow <.25cm> [.15,.25]    from  9 0  to 10 0

\plot  -6 -3    6 9    /
\arrow <.25cm> [.15,.25]    from  6 -6  to -7.07 7.07        
\arrow <.25cm> [.15,.25]    from  -6 -6  to 7.07 7.07        

\setlinear 

\put {\mathput{O}}              at 0.5 0

\put {\mathput{u}}              at -8 8
\put {\mathput{v}}              at 8 8 
\put {\mathput{z}}              at 10.5 -0.5
\put {\mathput{t}}              at -0.5 9.5

\put {\mathput{I}}                 at 4 2.5
\put {\mathput{II}}                at 4 5.5
\put {\mathput{III}}               at 4 8.5

\put {\mathput{u=0}}              at -7.3 -6
\put {\mathput{u=u_1}}              at -7.3 -3 

\setshadegrid span <1pt>
\vshade -6 -6 -3 <,z,,> 6 6 9 /
\endpicture}$$
\vspace{.5cm}
\caption{The null coordinate relationships in the $t$-$z$ plane (orthogonal to the plane wave fronts aligned with the $x$-$y$ planes) for a sandwiched radiation field between the two null hypersurfaces $u=0$ and $u=u_1>0$.
}
\label{fig:0}
\end{figure}

Fig. \ref{fig:0} illustrates the relationships between the coordinates for the case of {a radiation field sandwiched in a $u$-coordinate interval $[0,u_1]$}.

\subsection{The \lq\lq optical medium analogy"}

Working out the electromagnetic analogy, we find for the electric and magnetic permeability tensors (\ref{epsdef}) of the corresponding equivalent medium in the {wave-zone} the following expressions
\beq
\epsilon_{ab}=\mu_{ab}={\rm diag}[\epsilon_1,\epsilon_2,\epsilon_3]={\rm diag}\left[\frac{G}{F},\frac{F}{G},FG\right]\,, 
\eeq
while the rotation vector $M_a=0$ vanishes identically, so that the medium is linear and nongyrotropic. 
In the weak field  limit $b_{\rm(gw)} u \ll 1$ we have
\begin{eqnarray}
\epsilon_{1}&=&\frac{\cosh b_{\rm(gw)} u}{\cos b_{\rm(gw)} u}\sim 1+b_{\rm(gw)}^2 u^2 \,,\nonumber\\
\epsilon_{2}&=&\frac{1}{\epsilon_{1}}\sim 1-b_{\rm(gw)}^2 u^2\,,\nonumber \\
\epsilon_{3}&=&\cosh b_{\rm(gw)} u \cos b_{\rm(gw)} u\sim 1\,,
\end{eqnarray}
up to the second order.
The general definition (\ref{ndefstat}) of the refraction index then gives
\beq
n=\frac{FG}{\sqrt{G^2e_1^2+F^2e_2^2+F^2G^2e_3^2}}\,.
\eeq
The anisotropic properties of the medium are evident considering photons traveling in the three coordinate directions.
{For instance, for photons along the $x$-axis is ${\mathbf e}=\partial_x$ so that $n=n_x=F$, whereas for photons along the $y-$axis is $n=n_y=G$ and for photons along the $z-$axis is $n=n_z=1$.}
In fact, starting from the vanishing of the line element $ds^2=0$ for photons, we have
\begin{eqnarray}
dt^2&=&F(u)^2dx^2 +G(u)^2 dy^2 +dz^2\nonumber \\
&=& \epsilon_{2}\epsilon_{3} dx^2 +\epsilon_{1}\epsilon_{3} dy^2 + dz^2\,,
\end{eqnarray}
so that  the travel coordinate time of light can be expressed in terms of an associated optical path
\beq
dt = \sqrt{n_x^2 dx^2 +n_y^2 dy^2 + n_z^2 dz^2}\,,
\eeq 
whereas the identification
\beq
n_x=\sqrt{\epsilon_{2}\epsilon_{3}}\,,\qquad  n_y=\sqrt{\epsilon_{1}\epsilon_{3}}\,,\qquad n_z=1\,, 
\eeq 
characterizes the anisotropic properties of the medium.
Explicitly, in the wave region, we find
\beq
n_x=F(u)<1 \,,\qquad n_y=G(u)>1\,,
\eeq
and {hence it follows that the $x$- and $y$-axes are naturally defined as} the superluminal and subluminal direction of light propagation, respectively. It is worth noticing that such a definition holds only for coordinate components of the velocity of light in the chosen coordinate system.
 
In the in-zone we have simply
\beq
\epsilon_{1}=\epsilon_{2}=\epsilon_{3}=1\,,
\eeq
so that $n=n_x=n_y=n_z=1$, whereas in the out-zone {it is}
\beq
\epsilon_{1}=\frac{\gamma+\delta u}{\alpha+\beta u}=\frac{1}{\epsilon_{2}} \,,\quad
\epsilon_{3}=(\gamma+\delta u)(\alpha+\beta u)\,.
\eeq
In the latter case (out-zone) it is always possible to perform a coordinate transformation  {$(t,x,y,z)\to (T,X,Y,Z)$} that will convert the line element to the standard {Minkowskian} form
\begin{eqnarray}
\label{coordtrans}
U &=&u \,,\quad
X =F(u)x \,,\quad
Y =G(u)y \,,\nonumber \\ 
V&=&v+\frac12F(u)F'(u)x^{2}+\frac12G(u)G'(u)y^{2}\,,\nonumber \\
T&=& \frac{1}{\sqrt{2}}\left(V+U\right)\,,\quad
Z= \frac{1}{\sqrt{2}}\left(V-U\right)\,.
\end{eqnarray} 
In the new Cartesian coordinates, the electric permeability tensor is again diagonal and with unitary components.
Thus the refraction index $n$ in both in- and out-zone can be set equal to 1, {provided  the frame $\{\partial_U,\partial_V,\partial_X,\partial_Y\}$ is identified with $\{\partial_u,\partial_v,\partial_x,\partial_y\}$ in order to make the right comparison.}

\subsection{Wave propagation}

In the limit of geometric optics electromagnetic waves propagate along the null geodesics of the spacetime \cite{mas87}.
Therefore, one can study how an incoming light beam propagating along a straight line in the in-zone is deflected when passing through the scattering region before emerging in the out-zone, where it moves again along a straight line but with different direction.   

In the in-zone, the constant photon 4-momentum $P_{I}$ can be parametrized in terms of the conserved specific momenta $p_v$, $p_x$ and $p_y$
associated with the three Killing vectors $\partial_v, \partial_x,\partial_y$ as
\begin{eqnarray}\label{eq:U0}
P_{I} &=& -p_v\left(\partial_u+\frac{p_\perp^2}{2p_v^2}\partial_v\right) +p_\perp
\nonumber\\
  &=& -\frac{1}{\sqrt{2}}
 p_v \left(1+\frac{p_\perp^2}{2p_v^2}\right)\partial_t+p_\perp\nonumber\\
  &&   -\frac{p_v}{\sqrt{2}}\left( -1 +\frac{p_\perp^2}{2p_v^2}\right)\partial_z
\,,
\end{eqnarray}
where $p_\perp=p_x\partial_x+p_y\partial_y$ (with $p_\perp^2=p_x^2+p_y^2$ using the flat spacetime notation for convenience) and $p_v<0$ for $P_{I}$ to be future-pointing.

In the wave-zone we have instead
\begin{eqnarray}
\label{Pgeogw}
P_{II}&=&-p_v\partial_u-\frac{1}{2p_v}\left(\frac{p_x^2}{F(u)^2}+\frac{p_y^2}{G(u)^2}\right)\partial_v\nonumber\\
&&+\frac{p_x}{F(u)^2}\partial_x+\frac{p_y}{G(u)^2}\partial_y\,, \nonumber\\
&=&-\frac{p_v}{\sqrt{2}}
\left[ 1 + \frac{1}{2p_v^2}\left(\frac{p_x^2}{F(u)^2}+\frac{p_y^2}{G(u)^2}\right)\right]\partial_t\nonumber\\
&&-\frac{p_v}{\sqrt{2}}
\left[ -1 + \frac{1}{2p_v^2}\left(\frac{p_x^2}{F(u)^2}+\frac{p_y^2}{G(u)^2}\right)\right]\partial_z\nonumber\\
&&
+\frac{p_x}{F(u)^2}\partial_x+\frac{p_y}{G(u)^2}\partial_y
\,.
\end{eqnarray}

Finally, in the out-zone the constant photon 4-momentum in $(U,V,X,Y)$ coordinates is given by 
\beq
\label{Uregion3}
P_{III}= -Q_V\left(\partial_U+\frac{Q_\perp^2}{2Q_V^2}\partial_V\right) +Q_\perp\,,
\eeq
where $Q_\perp=Q_X\partial_X+Q_Y\partial_Y$ (with $Q_\perp^2=Q_x^2+Q_y^2$), and {the constants $Q_V,Q_X,Q_Y$ are determined} by imposing the matching conditions at the boundary II--III where $u = u_1$.
Thus one finds
\begin{eqnarray}
\label{IIImomenta2}
Q_V &=& p_v\,,\\ 
Q_X &=& p_v \sin (b_{\rm(gw)} u_1)b_{\rm(gw)} x_0 +p_x \cos(b_{\rm(gw)} u_1)\,,\nonumber\\ 
Q_Y &=& -p_v \sinh (b_{\rm(gw)} u_1)b_{\rm(gw)} y_0 +p_y \cosh(b_{\rm(gw)} u_1)\,,\nonumber
\end{eqnarray}
where $x_0$ and $y_0$ denote the initial values of the coordinates $x$ and $y$ {along the directions transverse  to the wave propagation.}
Recalling that at the boundary $u = u_1$ (i.e., the most relevant surface to study light scattering in terms of refraction phenomena) we have
{
\begin{eqnarray}
F(u_1)&=& \cos(b_{\rm(gw)} u_1)=n_x|_{u_1}\nonumber\\
G(u_1)&=& \cosh(b_{\rm(gw)} u_1)=n_y|_{u_1}\,,
\end{eqnarray}
}
and
{
\begin{eqnarray}
F'(u_1)=-b_{\rm(gw)}\sin(b_{\rm(gw)}u_1)&=&n_x'|_{u_1}\,,\nonumber\\ 
G'(u_1)=b_{\rm(gw)}\sinh(b_{\rm(gw)} u_1)&=&n_y'|_{u_1}\,,
\end{eqnarray}
}
(a prime denoting differentiation with respect to $u$), 
the transverse momentum in the out-zone can be written as
\beq
\label{Q_perp_gw}
Q_\perp= 
A
p_\perp -p_v  A' X_0\,,
\eeq
where 
\beq
A=\left(
\begin{array}{cc}
n_x & 0 \cr
0 & n_y
\end{array}
\right)\,,\qquad  A'=\left(
\begin{array}{cc}
n_x'  & 0 \cr
0 & n_y' 
\end{array}
\right)
\eeq
and $X_0=(x_0, y_0)$. The symbol $|_{u_1}$ has been omitted for simplicity.

\begin{figure}
\begin{center}
\includegraphics[scale=0.35]{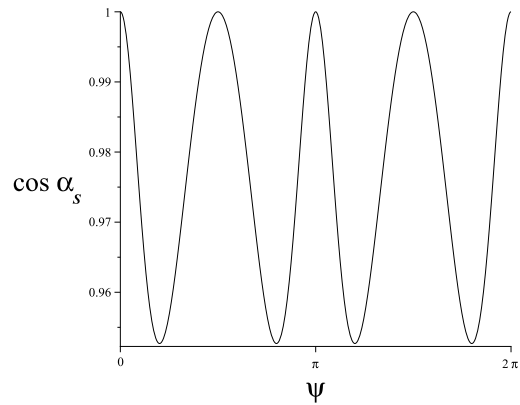}
\end{center}
\caption{[Gravitational wave case]  The angle $\alpha_{\rm s}$ given by Eq. (\ref{alpha_scatt}) and written as a function of the angle $\psi$ of the polar representation of transverse momenta in the case $b_{\rm (gw)} u_1=\pi/4$.
}
\label{fig:3}
\end{figure}

If the photon impacts the wave region (in the transverse plane) at the origin of the coordinates, i.e., $x_0=0=y_0$, Eq. (\ref{Q_perp_gw}) implies
\beq
Q_x=n_x p_x\,, \qquad Q_y=n_y p_y\,,
\eeq
{expressing along each axis the Snell law of refraction of ordinary optics.} 
Since $n_x\not=n_y$, the relation (\ref{Q_perp_gw}) for $X_0=0$ can be represented by an ellipse in the space of momenta, as shown in Fig. \ref{fig:1}.
One can also evaluate the angle between the directions of $p_\perp$ and $Q_\perp$, i.e., the scattering angle $\alpha_{\rm s}$, defined by
\beq
\label{alpha_scatt}
\cos \alpha_{\rm s} =\frac{p_\perp \cdot Q_\perp}{|p_\perp| |Q_\perp|}=\frac{n_xp_x^2+n_yp_y^2}{\sqrt{p_x^2+p_y^2}\sqrt{n_x^2p_x^2+n_y^2p_y^2}}\,.
\eeq
If $p_x=0$ (or $p_y=0$), one simply finds $\alpha_{\rm s}=0$.
Introducing the polar representation of transverse momenta $p_x=p_\perp \cos \psi$, $p_y=p_\perp \sin \psi$, the previous expression becomes
\beq
\label{alpha_scatt2}
\cos \alpha_{\rm s}=\frac{n_x\cos^2\psi+n_y\sin^2\psi}{\sqrt{n_x^2\cos^2\psi+n_y^2\sin^2\psi}}\,,
\eeq
which is invariant for $\psi\to\pi-\psi$.
Its behavior as a function of $\psi$ is shown in Fig. \ref{fig:3}. It turns out that the scattering angle $\alpha_{\rm s}$ remains close to zero in the whole range of allowed transverse momenta.

Vice versa, if the incident photon is purely longitudinal, 
i.e., $p_x=0=p_y$, Eq. (\ref{Q_perp_gw}) reduces to $Q_\perp=-p_v  A' X_0$.  In this case the photon travels along the same direction as the gravitational wave, a fact that makes trivial the whole  problem.

\begin{figure}
\begin{center}
\includegraphics[scale=0.3]{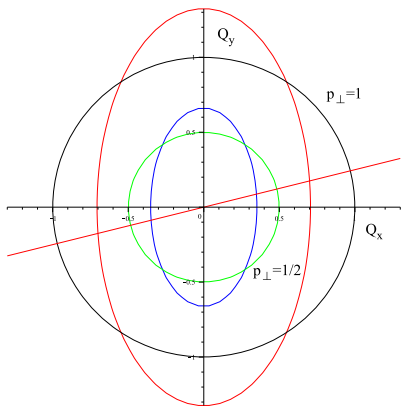}
\end{center}
\caption{[Gravitational wave case] In the space of transverse momenta, the different refraction indices $n_x$ and $n_y$ on the $x$ and $y$ axes imply that an initial circular distribution of momenta for incident photons becomes an ellipse  after the refraction process. The plot shows two cases: $p_\perp=1$ (external circle and ellipse) and  $p_\perp=1/2$ (internal circle and ellipse). The (generic) line intersecting the circle and the ellipse (in both cases) identifies corresponding  pairs $(Q_x,Q_y)$ and $(p_x,p_y)$ related by Eq. (\ref{Q_perp_gw}) with a choice of $X_0=0$ and using $b_{\rm (gw)} u_1=\pi/4$.
}
\label{fig:1}
\end{figure}

\section{Radiation scattering from a sandwiched electromagnetic wave}

Let us now consider  the similar situation in which an electromagnetic plane wave propagates along the positive $z$-axis and is sandwiched between two flat spacetime regions, as shown in Fig. (\ref{fig:1})  for the gravitational wave.
The corresponding conformally flat line element is given by Eq.~(\ref{lineelement}) with functions \cite{grif}
\beq
\label{FGem}
F=\cos(b_{\rm (em)}u)=G\,,
\eeq
differing from the corresponding gravitational wave case only by a trigonometric rather than hyperbolic cosine appearing in $G$, so that the above analysis is easily repeated.
The present case corresponds to a nonvacuum spacetime which is a solution of the Einstein equations with energy-momentum tensor
\beq
\label{Tmunu}
T=\phi_0 K \otimes K\,,\quad  K = {b_{\rm (em)}} \partial_v\,,
\eeq
where  $\phi_0=1/4\pi$ and the background quantity $b_{\rm(em)}$ is related to the frequency of the wave by $b_{\rm(em)}=\sqrt{2}\omega_{\rm(em)}$.

{As in the previous section, the  coordinate horizon of the metric at $ b_{\rm (em)}u=\pi/2$  is avoided by restricting the coordinate $u$ to the interval $[0,u_1]$ with $u_1<\pi/(2b_{\rm (em)})$.}
The matching conditions at the two null hypersurface boundaries now imply 
\begin{eqnarray}
\label{costanti2}
\alpha &=&\cos(b_{\rm (em)}u_1) + b_{\rm (em)}u_1\sin(b_{\rm (em)}u_1)=\gamma\,,\nonumber\\
\beta&=&-b_{\rm (em)}\sin(b_{\rm (em)}u_1)=\delta\,.
\end{eqnarray}

\begin{figure}
\begin{center}
\includegraphics[scale=0.3]{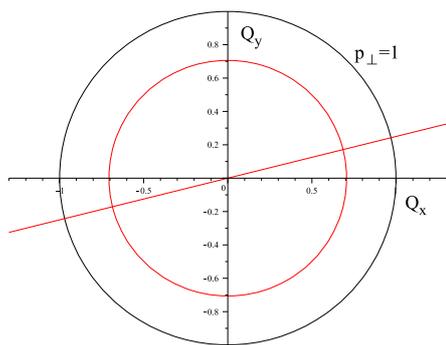}
\end{center}
\caption{[Electromagnetic wave case] In the space of transverse momenta, the coincident refraction indices $n_x=n_y$ on the $x$ and $y$ axes imply that an initial circular distribution of momenta for incident photons remains circular  after the refraction process. The plot shows the case  $p_\perp=1$ ($Q_\perp=1/\sqrt{2}$). The (generic) line intersecting the two circles  identifies corresponding  pairs $(Q_x,Q_y)$ and $(p_x,p_y)$ related by Eq. (\ref{Q_perp_em}) in the case $X_0=0$ and using $b_{\rm (em)} u_1=\pi/4$.
}
\label{fig:2}
\end{figure}

The previous analysis can be repeated straightforwardly, now implying in the wave region 
\beq
n_x=n_y=\cos b_{\rm (em)} u\equiv n\,,
\eeq
so that $n<1$.
The relation between {in- and out-momenta} is now given by
\begin{eqnarray}
\label{IIImomenta2em}
Q_V &=& p_v\,,\\ 
Q_X &=& p_v \sin (b_{\rm (em)} u_1)b_{\rm (em)} x_0 +p_x \cos(b_{\rm (em)}u_1)\,, \nonumber\\
Q_Y &=& p_v \sin (b_{\rm (em)} u_1)b_{\rm (em)} y_0 +p_y \cos(b_{\rm (em)} u_1)\,.\nonumber
\end{eqnarray}
Recalling that at the boundary
{
\begin{eqnarray}
F(u_1)&=& \cos(b_{\rm (em)}u_1)=n|_{u_1}\,, \nonumber\\
F'(u_1)&=& -b_{\rm (em)}\sin(b_{\rm (em)}u_1)=n'|_{u_1}\,,
\end{eqnarray}
}
the emerging transverse momentum can be written in the same form as in Eq. (\ref{Q_perp_gw}) with $A=nI$ and $A'=n'I$, $I$ being the identity matrix, so that 
\beq
\label{Q_perp_em}
Q_\perp=n p_\perp -p_v n'X_0\,.
\eeq
For $X_0=0$ this relation implies
\beq
Q_x=n p_x\,, \qquad Q_y=n p_y\,,
\eeq
which can be represented by a circle in the momentum space, as shown in Fig. \ref{fig:2}.
If $X_0=0$ (for simplicity) the scattering angle is always $\alpha_{\rm s}=0$, showing the difference between scattering by a gravitational and an electromagnetic wave.

\section{Concluding remarks}

{The optical medium analogy of a radiation field  sandwiched between two flat regions has been analyzed for the two cases where the radiation field is that
 of an exact gravitational plane wave or  that of an exact electromagnetic wave in a general relativistic context.}
In particular we have studied the typical scattering problem of a light beam by such a radiation field, elucidating the main features of the process in terms of associated optical properties of the equivalent optical active medium. 
The most relevant physics involves the  plane transverse to the direction of propagation of the wave: here 
the equivalent medium of the gravitational wave is inhomogeneous and anisotropic, whereas that of an electromagnetic wave is inhomogeneous but isotropic. In both cases the medium is active in the sense that the refraction index {depends} on time. Nevertheless, one can treat the scattering process as a series of multiple refractions through media with nearly constant refraction indices all the way up to the boundary of the interaction region, so that we are allowed to adopt the standard tools of ordinary optics.
  
Moreover, there exist directions in which the effective refraction index of the medium is {less} than 1 (so that the coordinate components  of the velocity of light become greater than 1). This realizes an interesting and novel  point of view for looking at the photon scattering by electromagnetic as well as gravitational waves in the exact theory of general relativity, which may also have a counterpart in experiments.  For instance, by means of  non-linear optical materials or  \lq\lq metamaterials" available nowadays \cite{Hiz,philbin}, one can arrange for an analogue material of the above mentioned radiation field spacetimes and compare the geometrization of physical interactions with experimental data.
In an experimental scenario, however, we are aware that a number of difficulties arises. For example, in order to mimic the transient event of a passing gravitational wave one would need to consider a moving optical medium. This can be remedied by constructing a medium that is susceptible to non-linear effects, such as an optical activation induced by an ultrashort, intense laser pulse. Then the optical medium itself remains at rest but its effective properties change as a propagating front when the medium is activated by the laser pulse. This method has been used previously to create fiber-optical analogs of a black hole's event horizon in laboratory \cite{philbin}.

\end{document}